\documentstyle[preprint,aps]{revtex}

\begin{document}
 \draft
\preprint{\small gr-qc/9402020 Alberta-Thy-48-93}

 \title
 {Self-dual gravity and the chiral model}
\author{Viqar Husain}
\address
 {Theoretical Physics Institute,\\
 University of  Alberta,
Edmonton, Alberta, Canada T6G 2J1}
\maketitle

\begin{abstract}
The  self-dual Einstein equation (SDE) is shown to be equivalent to
the two dimensional chiral model, with gauge group chosen as the
group of area preserving diffeomorphisms of a two dimensional
surface.
The approach given here leads to an analog of the Plebanski equations
 for general self-dual metrics, and to a natural Hamiltonian
formulation
 of the SDE, namely that of the chiral model.
\end{abstract}
\vfill
\eject

Since the pioneering works on the KdV equation, two dimensional
integrable models have been much studied and there a number of books
that discuss the developments \cite{fadtakh,das,newell}.

There are some basic features shared by all the models which provides
the clues to integrability. One of these is the existense of two
different Hamiltonian formulations for the same equation. This
feature allows a systematic way to construct the conserved quantities
and to prove that these are in involution. Another is the existence
of
the linear Lax form
of the  non-linear equations. This is basically the same as the two
dimensional zero curvature conditions that lead to the non-linear
equations. An example of this is the derivation of the KdV
equation from an SL(2,R) zero curvature condition using a particular
parametrization of the gauge field \cite{das}.

More recently it has been demonstrated that many integrable equations
are
also derivable from Yang-Mills self-duality conditions in four
dimensions.
For example, using  again various parametrizations of an  SL(2,R)
gauge
field, and this time imposing self-duality on the curvatures
followed by a
dimensional reduction to two dimensions, it has been shown that one
can
obtain the KdV, Sine-Gordon, and non-linear Schrodinger equations
\cite{ward,mason,bakas}. It has
also been shown that one dimensional reductions of the Yang-Mills
self-duality conditions lead to various known classical equations
depending on the choice of gauge group \cite{abl}.

The self-dual Einstein equation (SDE) is another much studied system.
Plebanski \cite{pleb} has given an elegant  formulation of these
equations
 in terms of one function of all the spacetime coordinates, referred
 to as the heavenly equations.
 There are indications that
this system is entirely integrable \cite{pen} although it has not yet
 been shown to have the standard features associated with
integrability
  that are mentioned above. There
 are a number of interesting results associated with these equations.
 Using a form of the SDE  \cite{ajs} suggested by the Ashtekar
Hamiltonian variables for general relativity \cite{ash}, a
connection with the self-dual Yang-Mills equation has been
demonstrated
\cite{masnew}:  the SDE  may be obtained from a 0+1  dimensional
reduction of  the self-dual Yang-Mills equation when the gauge group
is chosen as the  group of volume preserving diffeomorphisms of an
(auxilliary) three manifold. Another result is that  the field
equation for
the continuum limit  of the Toda model is the same as the SDE for a
special
 ansatz for the metric \cite{bakkir}. A further connection with
two dimensional theories has been the derivation of the Plebanski
equation \cite{pleb} for self-dual metrics from a large N limit of
the SU(N) chiral model \cite{park}.

In this letter it is shown how the chiral model field equations
may be derived in a simple and direct way from the (unreduced)
SDE. The starting point will be the relatively new way of
writing the SDE in a 3+1 form due to Ashtekar, Jacobson
and Smolin \cite{ajs}.

The chiral field $g(x,t)$ is a mapping from a 2d spacetime  into a
group $g$. The dynamics follows from the Lagrangian density
\begin{equation}
  L = {1\over 2}{\rm Tr}\bigl( \partial_\mu g^{-1}\partial_\nu
g\bigr)
\eta^{\mu\nu}
\end{equation}
where $\eta^{\mu\nu}$ is the flat Minkowski or Euclidean metric.
 The equations of motion are
\begin{equation}
 \partial_\mu (g^{-1}\partial_\mu g)=0
\end{equation}
If we define the Lie algebra valued 1-form  $A_\mu :=
g^{-1}\partial_\mu g$,
 then this equation of motion  becomes
 \begin{equation}
 \partial_\mu A_\mu =0.
 \end{equation}
 Since $A_\mu$ by
definition has a  pure gauge form, it follows that
\begin{equation}
F_{\mu\nu} = \partial_\mu A_\nu - \partial_\nu A_\mu + [A_\mu,A_\nu]
= 0.
\end{equation}
Associated with the gauge field $A_\mu$ there is also the covariant
derivative
 \begin{equation}
 D_\mu = \partial_\mu + A_\mu.
\end{equation}
 The chiral model describes flat connections $A_\mu$
satisfying  $\partial_\mu A_\mu = 0$. Eqns. (3-4) are the first order
forms of the field equation (2).

 The SDE can also be written in a first order form using
the Ashtekar Hamiltonian variables for general relativity \cite{ajs}.
Self-duality is the essential ingredient for this canonical
formulation
 and it is natural to ask how the SDE looks in it.
The phase space coordinate is the spatial projection
of the (anti)self-dual part of the spin connection and its conjugate
momentum  is a densitized dreibein.  The same is  true for
Euclidean or (2,2) signatures, or  complex general relativity, which
are
the cases of interest  for self-dual Riemann curvatures.

In these Hamiltonian  variables, we would like to know  what is the
phase
space condition  corresponding to the vanishing of the
(anti)self-dual part
of the four dimensional Riemann curvature. The answer is that the
spatial
 projection of the latter  must be zero.
 The vanishing of this spatial projection, when  substituted into
 Ashtekar's 3+1 evolution equations leads to the new form of the SDE.
 It is straightforward to verify that
 this condition  remains zero under the Hamiltonian evolution.
 The resulting  equations on four-manifolds $M=\Sigma^3{\cal T}imes
R$ may be
 written in terms of three spatial vector fields $V_i^a$ on
$\Sigma^3$:
\begin{eqnarray}
 Div V^a_i &=& 0  \\
 {\partial V_i^a\over \partial t} &=& {1\over 2} \epsilon_{ijk}[
V_j,V_k]^a,
 \end{eqnarray}
where the divergence is defined with respect to a constant  auxiliary
density
 and the right hand side of (7) is the Lie bracket.
The self-dual
four metrics  are constructed from  solutions of these equations
using
\begin{equation}
g^{ab} = ({\rm det} V)^{-1}  [ V^a_i V^b_j \delta^{ij}  + V_0^a V_0^b
].
\end{equation}
Here  $i,j,k...=1,2,3$ label the vector field, $a,b,...$ are abstract
vector
indices,
$V_0^a$ is the vector field that is used to perform the 3+1
decomposition,
and
$ \partial V_i^a/\partial t \equiv V_0^b\partial_b V_i^a$.
 The time derivative in (7) can be written in the more general form
 $[V_0,V_i]^a$.
(For details of the derivation of these equations the reader is
referred  to \cite{ajs} where they were originally derived, or the
review in \cite{vh}).

The starting point will be the SDE in the form (6-7).
We first rewrite equation (7) in a form similar to that suggested by
Yang \cite{yang} for the self-dual Yang-Mills equation
 \begin{equation}
 F_{ab}={1\over 2}\epsilon_{ab}^{\ \ cd}F_{cd}
 \end{equation}
  on a complex manifold. Replacing the (local) complex flat
coordinates
 $x_0,...,x_3$ by the linear combinations $t=x_0+ix_1$,
 $u=x_0-ix_1$, $x=x_2-ix_3$ and  $v=x_2+ix_3$,
 equation (9) becomes
 \begin{eqnarray}
  F_{tx}& = &F_{uv}=0 \\
 F_{tu}& + &F_{xv}=0.
 \end{eqnarray}

For the SDE, defining in a similar way
\begin{eqnarray}
{\cal T} &=& V_0+iV_1 \ \ \ \ \ \ \ \ \ {\cal U} = V_0-iV_1 \nonumber
\\
{\cal X} &=& V_2-iV_3 \ \ \ \ \ \ \ \ \ {\cal V}=V_2+iV_3,
\end{eqnarray}
the evolution equations (7) become
\begin{equation}
[{\cal T},{\cal X}] = [{\cal U},{\cal V}]=0
\end{equation}
\begin{equation}
 [{\cal T},{\cal U} ] + [{\cal X},{\cal V}]=0,
\end{equation}
where the vector indices have been suppressed. This shows a rather
direct
analogy between  the self-dual Yang-Mills and Einstein equations,
namely,
the Yang-Mills curvatures in equations (10-11) are replaced by the
Lie
brackets of the vector fields. (This analogy has been noted in a
related
way in ref. \cite{masnew}, and equations (13-14) have also been
studied in
ref. \cite{cmn} from a different viewpoint to that below.)

We now show how the SDE may be written as a chiral model
field equation. Fixing a local coordinate system $t,x,p,q$,
the volume form is $\omega = dt\wedge dx\wedge dp \wedge dq$, with
respect to which we define the divergence in equation (6).
 We take now the following divergence free form

for the vector fields ${\cal T},{\cal X},{\cal U},{\cal V}$ in terms
of

two functions,
$A_0(t,x,p,q)$
 and $A_1(t,x,p,q)$:
 \begin{eqnarray}
 {\cal T}^a &=& ({\partial \over \partial t})^a \nonumber \\
 {\cal X}^a &=& ({\partial \over \partial x})^a  \nonumber \\
 {\cal U}^a &=& ({\partial \over \partial t})^a +

 \alpha^{ba}\partial_b A_0
 \nonumber \\
 {\cal V}^a &=& ({\partial \over \partial x})^a +

 \alpha^{ba}\partial_b A_1,
 \end{eqnarray}
 where $\alpha^{ab} = (\partial/\partial p)^{[a}\otimes
 (\partial/\partial q)^{b]}$  is the antisymmetric tensor that is the
 inverse of the two form $(dp\wedge dq)_{ab}$ in the $p,q$ plane.
 (This form for the vector fields is similar to, but more general
than that
 used previously by the author in ref. \cite{vh}, where one-Killing
field
 reductions of the  self-duality equations are discussed.)
 Substituting equations (15) into (13-14) gives
 \begin{eqnarray}
 \alpha^{ab}\partial_b \bigl[\partial_0 A_1 - \partial_1 A_0 +
 \{A_0,A_1\}\bigr] &=& 0\\
 \alpha^{ab}\partial_b \big[\partial_0A_0 + \partial_1 A_1\bigr] = 0.
 \end{eqnarray}
 where the bracket on the left hand side of equation (16) is the
 Poisson bracket with respect to $\alpha^{ab}$:
 \begin{equation}
 \{A_0,A_1\} := \alpha^{ab}\partial_aA_0\partial_bA_1=
 \partial_p A_0 \partial_q A_1
 -\partial_q A_0 \partial_p A_1,
 \end{equation}
 and $\partial_0,\partial_1$ denote partial derivatives with respect
to
 $t,x$ etc. Equations (16-17) imply that the terms in their
 square brackets are equal to two arbitrary functions of $t,x$, which
 we write as
 \begin{eqnarray}
  \partial_0 A_1 - \partial_1 A_0 +
 \{A_0,A_1\} &=& \partial_0 F(x,t) +\partial_1 G(x,t)\\
 \partial_0A_0 + \partial_1 A_1 &=& \partial_1 F(x,t) - \partial_0
G(x,t),
\end{eqnarray}
(where $F(x,t),G(x,t)$ are arbitrary.)
With the redefinitions
\begin{equation}
a_0(t,x,p,q):= A_0 + G \ \ \ \ \ \ a_1(t,x,p,q) := A_1 - F,
\end{equation}
equations (19-20) become
\begin{eqnarray}
\partial_0 a_1 - \partial_1 a_0 +
 \{a_0,a_1\} &=& 0 \\
 \partial_0 a_0 + \partial_1 a_1 &=& 0.
 \end{eqnarray}
 These are precisely the chiral model equations (3-4) on the
 $x,t$ `spacetime', with $p,q$ treated as coordinates on
 an `internal' space, and with the commutator in (4) replaced by the
 Poisson bracket with respect to $\alpha^{ab}$.
 The gauge group is therefore the  group of transformations that
preserve
 $\alpha^{ab}$ on the internal $p,q$ space. (Note that the
 redefinitions (21) do not alter the vector fields ${\cal U},{\cal
V}$ in
 equations (15).)

 There is an important  question regarding the relation between the
full SDE (7), and the chiral model equations (22-23)
derived from them. How general is the form (15) for the vector
fields?  We can see that the two first order equations for $a_0,a_1$
are equivalent to a single second order equation for a function
$\Lambda(t,x,p,q)$. Equation (23) implies
 \begin{equation}
 a_0=\partial_1\Lambda  \ \ \ \ \ \ \
 a_1=-\partial_0\Lambda.
 \end{equation}
 With this the  vector fields ${\cal U},{\cal V}$ in equation (15)
become
\begin{eqnarray}
{\cal U} &=&  {\partial\over \partial t} -
 \Lambda_{xq}{\partial \over \partial p} +
 \Lambda_{xp}{\partial \over \partial q} \nonumber \\
{\cal V} &=& {\partial\over \partial x}+
\Lambda_{tq}{\partial \over \partial p} -
 \Lambda_{tp}{\partial \over \partial q}.
\end{eqnarray}
and equation (22) becomes
\begin{equation}
\Lambda_{tt} + \Lambda_{xx} +
\Lambda_{xp} \Lambda_{tq} - \Lambda_{xq} \Lambda_{tp}=0,
\end{equation}
(where the subscripts denote partial  derivatives.)
 Equation (26) is one equation for a function of all the spacetime

 coordinates
and therefore doesn't represent any reduction in the local degrees of
freedom for self-dual metrics. Using equation (8) the line element is
\begin{eqnarray}
ds^2 &=&
    dt ( \Lambda_{tp}dp + \Lambda_{tq}dq )
 + dx ( \Lambda_{xp}dp + \Lambda_{xq}dq )\nonumber \\
 & & - {1\over \{\Lambda_t,\Lambda_x\}}\bigl(
     ( \Lambda_{xp}dp + \Lambda_{xq}dq )^2
    + ( \Lambda_{tp}dp + \Lambda_{tq}dq )^2\bigr)
\end{eqnarray}

For comparison, and to see the generality of the
form of the vector fields used in equation (25), we note how
Plebanski's

first heavenly equation for general self-dual metrics may be derived
from

equations (13-14) \cite{cmn}. Working again in specific coordinates
and
taking the same form  for ${\cal T},{\cal X}$ as in equation (15)

(which solves the
first  of equations (13)), let, for some function
$\Omega(t,x,p,q)$
 \begin{eqnarray}
 {\cal U} &=& -\Omega_{xq}{\partial \over \partial p} +
 \Omega_{xp}{\partial \over \partial q} \nonumber \\
 {\cal V} &=& \Omega_{tq}{\partial \over \partial p} -
\Omega_{tp}{\partial \over \partial q}.
\end{eqnarray}
The vector fields (28) solve equation (14), while the second equation
in (13) leads (after a few steps) to Plebanski's
first equation
\begin{equation}
\Omega_{xp}\Omega_{tq} - \Omega_{xq}\Omega_{tp} = 1.
 \end{equation}
A comparison of equations (25) and (28) shows the vector fields
${\cal U},{\cal V}$  in each equation have the
same functional content. Therefore equation (26) is an alternative
to Plebanski's equation (29).

An advantage of equation (26) over the Plebanski one (29) is that the
former  has a natural Hamiltonian formulation which is just that of
the
chiral model. This Hamiltonian formulation is given in, for example,
ref.\cite{fadtakh} for finite dimensional groups, and its
generalization
 to the infinite dimensional case of relevance here is immediate.

There is now also the possibility of approaching the SDE directly
from the two dimensional model point of view, and investigating
integrability using the standard methods. For example, one can derive
conservation laws \cite{vh2} via this approach, and ask if
there is a second Hamiltonian formulation just as for other
integrable models.

However some remarks are in order regarding this because

global spacetime considerations need to be addressed before a

Hamiltonian can be written down. The internal gauge group

for the chiral model

must first be fixed to be  the group of area preserving
diffeomorphisms

of a {\it specific} two dimensional surface. This fixes part of the

topology of the self-dual manifold. The topology of the
two-dimensional

chiral model background remains to be specified. From this viewpoint

therefore, there is not one but an infinite number of Hamiltonian

formulations specified by the gauge group and the chiral model

background, with each phase space associated with a particular sector
of

self-dual metrics. A further question in this regard is how

large a class of solutions to the SDE results from a given

internal group and chiral model background. Whereas the chiral model

solution space is infinite dimensional, the metrics derived from the

solutions may be related by diffeomorphisms since the coordinates
have

been only partially fixed in (15). In summary, care is needed in
making

 statements of a global nature given that all derivations in this

paper involve local considerations.

Investigating the quantum theory via canonical quantization may also
be
of interest since the SDE constitute the largest midi-superspace
model.
The existence of an infinite number of conservation laws for this
system \cite{vh2,strach,grant}, unlike the full Einstein equations
\cite{tor}, leads to the possibility of an infinite
number of fully gauge invariant classical observables to represent as
linear operators on the Hilbert space of the theory. A proper
quantization should lead to a description of quantum
 `non-linear gravitons'\cite{pen}.

In summary, we have shown how the chiral model field
equations can represent the full SDE starting from the
Ashtekar-Jacobson-Smolin form of the latter. This result also
gives an alternative to the Plebanski equations.

This work was supported by the Natural Science and Engineering

Research Council of Canada.

\end{document}